\documentclass[apjl]{emulateapj}

\usepackage[usenames]{color}

\usepackage{ulem}
\newcommand{\be}{\begin{equation}}
\newcommand{\ee}{\end{equation}}
\newcommand{\bea}{\begin{eqnarray}}
\newcommand{\eea}{\end{eqnarray}}

\newcommand{\mpci}{\mbox{$h$\,Mpc$^{-1}$}}
\newcommand{\thrd}{three-dimensional}
\newcommand{\twod}{two-dimensional}

\begin{document}

\title{Exploring Large-scale Structure with Billions of Galaxies}
\author{Hu Zhan, Lloyd Knox, J.~Anthony Tyson, and Vera Margoniner} 
\shortauthors{Zhan et al.}
\shorttitle{Exploring Large-scale Structure}
\affil{Department of Physics, University of California, Davis, CA 95616}
\email{zhan@physics.ucdavis.edu}
\email{lknox@physics.ucdavis.edu}
\email{tyson@physics.ucdavis.edu}
\email{vem@physics.ucdavis.edu}

\begin{abstract}
We consider cosmological applications of galaxy number
density correlations to be inferred from future deep and wide
multi-band optical  surveys.  We mostly focus on very
large scales as a probe of possible features in the primordial power
spectrum.  We find the proposed  survey of the Large
Synoptic Survey Telescope may be competitive with future
all-sky CMB experiments over a broad range of scales.  On very large
scales the inferred power spectrum is robust to photometric redshift
errors, and, given a sufficient number density of galaxies, to angular
variations in dust extinction and photometric calibration errors.  We
also consider other applications, such as constraining dark energy
with the two CMB-calibrated standard rulers in the matter power
spectrum, and controlling the effect of photometric redshift errors to
facilitate the interpretation of cosmic shear data. We find that
deep photometric surveys over wide area can provide constraints that
are competitive with spectroscopic surveys in small volumes.
\end{abstract}


\section{Introduction} \label{sec:intr}
Results from the \emph{Wilkinson Microwave Anisotropy Probe} 
\citep[\emph{WMAP},][]{bhh03} have revealed a CMB temperature power 
spectrum that is remarkably well-fit by a simple 5 or 6-parameter model
\citep[e.g.][]{svp03}.  The success of this simple model
is somewhat qualified by irregularities on large scales.  There is
evidence for departures from statistical isotropy, an anomalously low
quadrupole amplitude, 
an anomalous absence of correlations on angular scales
greater than 60 degrees, non-Gaussianity, and sharp features in the
temperature power spectrum
\citep{lwr03, peiris03, deoliveira04, efstathiou04b, eriksen04b, 
hansen04, schwarz04, land05, jaffe05}.  
While these irregularities may eventually
be revealed to be due to improper modeling of the instrument or
astrophysical foregrounds, they certainly have served to call
attention to the possibility of interesting departures from the
standard paradigm on the largest observable scales.  In this paper
we explore how well these scales can be probed by forthcoming
deep and wide photometric redshift surveys such as the  
survey of the Large Synoptic Survey 
Telescope\footnote{\url{see http://www.lsst.org.}} (LSST).

It has been suggested that by combining CMB observations and 
the Sloan Digital Sky Survey (SDSS) one can probe very large scales 
and thus detect possible features in the primordial power spectrum
 \citep*{wang99}.
We find that very large-volume photometric redshift surveys can
probe these large scales with smaller statistical errors.  
For a sufficiently large volume the galaxy
power spectra not only provide a complementary look at structure on
very large scales, but they can also determine the power spectrum on a
given scale even more precisely than with CMB temperature
anisotropy measurements.  The
reason is that in determining a power spectrum one is fundamentally
limited by the number of modes on a given scale, and there are more
such modes with a three-dimensional survey than with a two-dimensional
survey that encloses the three-dimensional survey.

Determining the power spectrum of the fluctuations on very
large scales is of great interest for several reasons.  First and
foremost, the power spectrum on these scales can be cleanly used
to infer the power spectrum of primordial fluctuations, and the
primordial fluctuation power spectrum on any scale is of interest.
The primordial power spectrum is one of our only handles on the mechanism
that led to the fluctuations that are responsible for the diversity
of structures we see in the Universe today, including ourselves.  

In the context of inflation, probing larger scales means probing
inflation at an earlier epoch.  Larger scales exited the fixed-size 
horizon earlier and thus had more time
to expand.  Reaching back to these earlier epochs may provide us
with valuable clues about inflation.

Finally, exploring these large scales is perhaps the best way, other
than precision measurement of CMB polarization on large angular scales,
to test the claims of statistical anisotropy in the \emph{WMAP} 
temperature maps.

Others have suggested ways to follow up on the large-scale \emph{WMAP} 
irregularities. To shed further light on the largest-scale
irregularity of all, the low quadrupole amplitude, \citet*{dhl04} 
and \citet{ss04} have proposed to use polarization data.
\citet*{kkc03} have proposed to use all-sky cosmic shear
measurements.  We must also note that the low quadrupole on the sky is
actually not that unlikely in a $\Lambda$CDM model, and could simply
be a statistical fluctuation (\citealt{blw03}; \citealt*{ccl03}; 
\citealt{cpk03}; \citealt{gwm03}; \citealt{oew04}).  
More conservative foreground 
modeling also reduces the level of discrepancy \citep{ebg04,slosar04}.
There are also statistical analyses with results that are consistent 
with statistical isotropy \citep*{hajian05}.

When independent measurements of the cosmic shear and galaxy power 
spectrum are combined, one can obtain robust constraints on dark energy
\citep{hj04}.
Furthermore, \citet{p04} showed explicitly that with a determination
of the galaxy power spectra as a function of redshift, one can then
indirectly infer the cosmic shear power spectrum with a lower 
statistical uncertainty than one can directly from the shear maps.  
While the 
LSST survey will sharply address dark energy via cosmic shear tomography
of billions of galaxies, here we emphasize using galaxy photometry (not
shear) in that survey to determine the matter power spectrum.

Deep and wide surveys are not only critical for measuring the power
spectrum on very large scales, they also lead to reduced sample
variance errors for power spectrum measurements on smaller scales.
We therefore consider applications of intermediate-scale constraints
on the power spectrum as well.  Namely, 
the broadband shape and baryon acoustic oscillations 
\citep{py70, be84, h89, hs96} in the matter 
power spectrum can serve as standard rulers to determine the angular 
diameter distance $r(z)$ and to constrain cosmological parameters
(\citealt*{eht98}; \citealt{cooray01}; \citealt{hh03}; \citealt{l03}; 
\citealt{se03}; \citealt{matsubara04}). We find the 
LSST survey capable of $\sim 1\%$ distance measurements to redshifts 
between 0.2 and 3.

The paper is organized as follows. Section \ref{sec:survey} 
describes our fiducial survey: the proposed LSST  survey.
The errors on the matter power spectrum are forecast 
in Section \ref{sec:error}, where effects 
of photometric redshift errors, redshift distortion, and spherical 
survey geometry are illustrated. We demonstrate in Section 
\ref{sec:extin} that galaxy counts from the fiducial survey can be 
used to determine the differential 
extinction and photometry errors so that these errors do not 
significantly contaminate the power spectrum. Section \ref{sec:app}
shows that comparable precision on the matter power spectrum can be 
achieved by the LSST and CMB, and that one can measure 
angular-diameter distances to percent level precision from baryon 
acoustic oscillations with the LSST. We conclude in Section 
\ref{sec:dis} with remarks on several challenges to measuring the 
matter power spectrum from photometric galaxy redshift surveys. A
brief summary of the spherical harmonic analysis, which is 
convenient for the survey geometry, is given in the Appendix.

\section{Fiducial Survey} \label{sec:survey}
The LSST will image 23,000 square degrees at high
galactic latitude deeply in 5--6 wavelength bands from 0.3--1.1 microns
($grizy$ system). Each band will have several hundred 15-second
exposures. 
Given its wide coverage and survey depth, the LSST is ideal for
measuring the galaxy power spectrum on very large scales.  To estimate
the galaxy redshift distribution for the 23,000 square degree survey
we degrade the Hubble Deep Field North (HDF-N), which has 
well-measured redshifts, to match LSST depth and image quality.

We approximate LSST's filter system using deep HST imaging and 
ground-based $J$-band ($\lambda_{\rm central} \sim 13000$\,\AA)
imaging from the 4-m KPNO telescope.
First, we convolve the HDF-N UBVI space images \citep{wbd96} with a
0.8" FWHM Gaussian to simulate the worst case seeing conditions.  
Then, we re-pixelize, add noise, and catalog the images to match the 
expected data quality for the final full-depth stack of 
$500 \times 15$s coadded exposures. For example, the final stack will 
go to 26.7 mag and 25.4 mag (10 and 30$\sigma$) in the $i$ band. 
The $J$-band image is left unchanged, because it has the same
resolution and worse seeing than expected for LSST.

We then compute photometric redshifts for all detected objects using a
technique based on SED fitting, and on a magnitude prior (Margoniner
2005, in preparation). The density of objects is $\sim 250,000$ per 
square degree. Because we need galaxies with well-determined redshifts,
we choose to keep only the ones with a sharp and well-defined peak in 
its redshift probability distribution; i.e., we require that the width 
of the peak be less than $0.04(1+z)$.
The resulting rms photometric redshift error is somewhat 
larger than this threshold width of the redshift peak. 
Our final object density after this cut
is $\sim 130,000$ per square degree, with a redshift distribution 
$\bar{n}_{\rm g}(z)$ parameterized as:
\be
\bar{n}_{\rm g}(z)= 530 z^2e^{-z/0.32} \mbox{ arcmin}^{-2}.
\ee

This exercise results in an rms of $\sigma_z = 0.065(1+z)$.  We take
this as a conservative upper limit to the rms and adopt
$\sigma_z = 0.04(1+z)$ as our fiducial value, which may be achievable
by further color-based cuts or priors, although this has yet to be 
demonstrated.Other surveys have achieved similar accuracy.  
For example, with luminous
red galaxies from the SDSS, \citet{pbs05}
find that $\sigma_z \sim 0.03$ for $z < 0.55$ and $\sim 0.06$ for 
$z < 0.7$. We also assume that the photometric redshift bias is
calibratable to 0.01(1+z) so that it does not significantly 
degrade the reconstruction of the matter power spectrum\footnote{
To roughly assess the effect of the redshift bias error, one may 
substitute the bias with a fractional distance error of the same 
magnitude. As shown in Fig.~\ref{fig:pk}, a distance error of $3\%$
(caused by an error in the matter fraction $\Omega_{\rm m}$) does 
not degrade the reconstruction much. Hence, a redshift bias of 
$1\%$ or less will be tolerable.}. 
This bias error is actually less stringent than what is required for
precise measurements of dark energy equation of state parameters 
through weak lensing or baryon acoustic oscillations.
 
\section{Measuring the Matter Power Spectrum} \label{sec:error}

On large scales, the cosmic density field can
be approximated by a Gaussian random field. At redshift $z$, the 
variance of the Fourier transform of the galaxy
number density contrast is
\be \label{eq:defPk}
\langle \delta_{\rm g}(\mathbf{k})\delta_{\rm g}^*(\mathbf{k})\rangle 
= P_{\rm g}(\mathbf{k}) = b^2(z) g^2(z) P(k) + \bar{n}_{\rm g}^{-1},
\ee
where $P_{\rm g}(\mathbf{k})$ is the galaxy power spectrum, $P(k)$ the 
matter power spectrum at $z = 0$, $b(z)$ the galaxy bias, and $g(z)$ 
the growth factor. We have ignored redshift distortion in equation 
(\ref{eq:defPk}) for the moment. If the 
factors multiplying $P(k)$ in the above equation are known, each 
Fourier mode of the density contrast can be used to make a very rough 
estimate of $P(k)$.

An unknown scale-dependent and stochastic bias will limit our ability 
to determine the matter power spectrum. However, the galaxy bias is 
thought to be scale-independent and deterministic on large scales.  
This expectation has been confirmed for 
$0.02\,\mpci{} < k < 0.1 \,\mpci{}$
from the power spectrum analysis of the SDSS
galaxies \citep{tbs04}, although the bias increases with 
luminosity. Extensive studies with the halo model, weak lensing,
and simulations \citep{ps00, hvg02, hj04, wdk04, smm05} will help us 
better understand the limits of galaxy bias. Here, we assume that the 
bias is known and scale-independent.

\subsection{Procedures}

Measuring the \thrd{} power spectrum on giga-parsec scales brings 
about two unique issues. 

First, one needs the comoving distance $r(z)$ and comoving angular 
diameter distance to convert the galaxy number density in 
redshift and angular coordinates to that in distance coordinates.
We assume hereafter that the curvature is known so that only one 
distance is needed.
For a shallow survey at low redshift, Hubble's law suffices to 
provide an approximation of $r(z)$, but for a survey that reaches 
$z = 2.5$ one must assume a cosmological model to calculate $r(z)$, 
or, possibly, obtain it from baryon acoustic oscillations.

Second, because galaxies are observed on 
a very long light-cone, there is considerable evolution of the bias 
and density fluctuations within the survey. If not accounted for, 
this evolution will contaminate the inferred power spectrum on 
scales (along the light-cone) over which the bias or growth 
factor changes appreciably.

For a photometric survey, there is yet another issue, namely, the
photometric redshift error. It suppresses the power spectrum in a 
similar way as the pairwise velocity dispersion but with greater 
magnitude. 

\begin{figure}
\centering
\epsscale{0.8}
\plotone{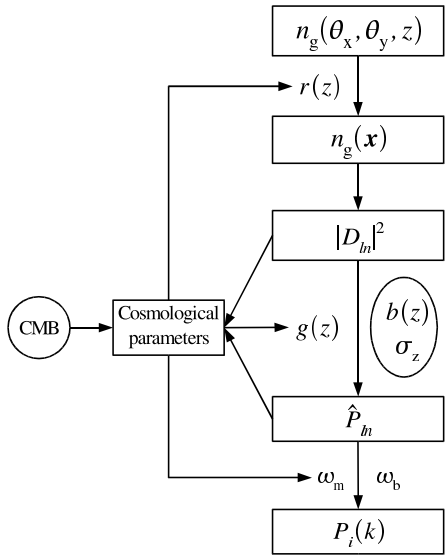}
\caption[f1]{Procedures for measuring the large-scale matter
power spectrum with the LSST  survey.
\label{fig:flow}}
\end{figure}

To summarize, we outline in Fig.~\ref{fig:flow} the procedures of 
measuring the matter power spectrum with the LSST  
survey. The survey provides the galaxy number density 
$n_{\rm g}(\theta_{\rm x},\theta_{\rm y},z)$, which is converted 
to $n_{\rm g}(\mathbf{x})$ with the knowledge of $r(z)$ from a 
specific cosmological model determined by the CMB or from baryon 
oscillations. The raw \thrd{} 
galaxy power spectrum $|D_{ln}|^2$ can then be calculated. The
subscripts $ln$ are the result of the spherical harmonic analysis
(see the Appendix). This raw power spectrum is strongly suppressed
in the radial direction. With a proper modeling of the effect of 
photometric redshift errors, redshift distortions, and galaxy bias,
one can estimate cosmological parameters \citep*[e.g.][]{cole94,
ballinger96, matsubara96, matsubara01}. After the above correction,
one obtains an estimate of the \thrd{} matter power spectrum 
$\hat{P}_{ln}$. Cosmological parameters can also be estimated from 
$\hat{P}_{ln}$ through 
the features imprinted in it and be fed back to the beginning of the
procedures. Finally, the primordial matter power spectrum 
$P_{\rm i}(k)$ can be reconstructed from $\hat{P}_{ln}$ with inputs 
of the matter density $\omega_{\rm m}$ and baryon density 
$\omega_{\rm b}$ and compared with that from the CMB 
\citep[e.g.][]{blw03}.

\begin{figure}
\centering
\epsscale{1}
\plotone{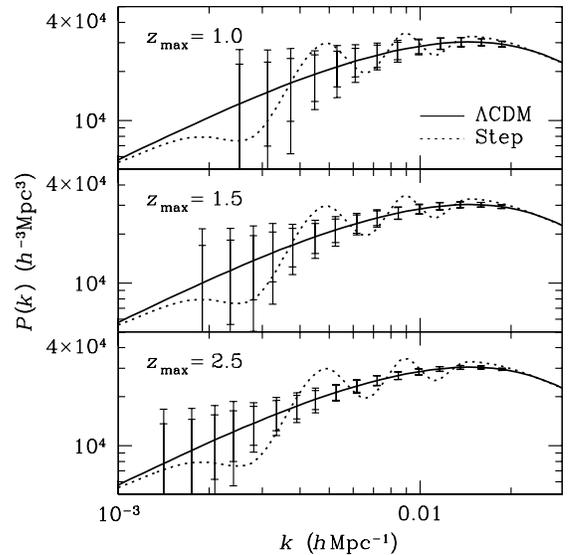}
\caption[f2]{Effect of depth ($z_{\rm max}$) on error forecasts for 
measurements of the matter power spectrum with the LSST. 
A 23,000 square degree photometric galaxy 
redshift survey is assumed. The solid
line is the fiducial model power spectrum, while the dotted line is 
the power spectrum generated by the step inflation potential
\citep{peiris03}. The error bars are 1$\sigma$ 
statistical errors of the power spectrum 
measured in non-overlapping logarithmic bins with 
bin width $\Delta k \simeq 0.16 k$. The inner
error bars are based on the simple mode-counting in a cubic volume,
e.g.~equation (\ref{eq:Nk}), while the outer ones count spherical 
harmonic modes using equation (\ref{eq:NkSHA}). All the power spectra
are scaled to $z = 0$.
\label{fig:varP}}
\end{figure}

\subsection{Forecasting Errors on $P(k)$}

For simplicity, and because we are only interested 
in the largest scales, we drop the negligible shot-noise term, as
justified below. 
Thus, the standard error in $P(k)$ estimated from modes within a band of 
width $\Delta k$ is simply
\be \label{eq:sigmaP}
\sigma_P(k)=\sqrt{2/N_k}P(k),
\ee
where $N_k$ is the number of independent modes in a shell of width
$\Delta k$ centered at $k$. 
For a cubic survey with volume $V$, $N_k$ is given by
\be \label{eq:Nk}
N_k =  k^2 \Delta k V / 2\pi^2.
\ee
Variants of these equations can be found in 
\citet*{fkp94} and \citet{t97}. We will see 
that this simple description of the errors is a good approximation.

To probe the very largest length scales, we want to go as deep as 
possible.
As an example, we show in Fig.~\ref{fig:varP} the power spectrum of a 
concordance model consistent with the \emph{WMAP} \citep{svp03} 
along with the power spectrum generated by a step inflation potential 
\citep{peiris03}.  This latter power spectrum provides a better fit to the
\emph{WMAP} data. If we cut the fiducial survey at 
$z_{\rm max}=1.0$, then the sample variance in the band power is too 
large to detect or rule out the step potential. 
However, it becomes possible when $z_{\rm max}$ extends to $2.5$.

Our description of the errors on $P(k)$ assumes a cubic geometry 
without redshift distortions or photometric redshift errors.  In
reality, the survey volumes are wedges from a sphere, with redshift
distortions and errors in the radial direction. Spherical harmonic 
analysis (for details, see the Appendix) is well suited for the real 
situation. A
simple estimator of the matter power spectrum $\hat{P}_{ln}$ can be 
constructed from spherical harmonic modes $D_{lmn}$ of the observed 
\thrd{} galaxy distribution:
\be
\hat{P}_{ln} = \frac{1}{E_{ln}} \left(\frac{1}{2l+1}
\sum_m |D_{lmn}|^2 - N_{ln} \right),
\ee
where $l$ and $m$ enumerate the usual angular modes of the spherical 
harmonics $Y_{lm}(\hat{\mathbf{r}})$, $n$ is associated with the radial 
modes $k_{ln}$ of the spherical Bessel function 
$j_l(k_{ln}r)$, and
$N_{ln}$ is the shot noise. The ensemble 
average of the estimator is
\be \label{eq:pln}
\langle \hat{P}_{ln} \rangle = 
\frac{1}{E_{ln}}\left(\langle |D_{lmn}|^2\rangle - N_{ln}\right) 
= \frac{1}{E_{ln}}\sum_{n'} U_{lnn'}^2 P(k_{ln'}),
\ee
where the ensemble average $\langle |D_{lmn}|^2\rangle$ is 
independent of $m$,
$U_{lnn'}^2$ accounts for photometric redshift errors and the 
linear redshift distortion, and $E_{ln} = \sum_{n'} U_{lnn'}^2$. 
We neglect the nonlinear redshift distortion because it behaves in a 
similar way as the photometric redshift error but has a much smaller 
effect.

\begin{figure}
\centering
\epsscale{1}
\plotone{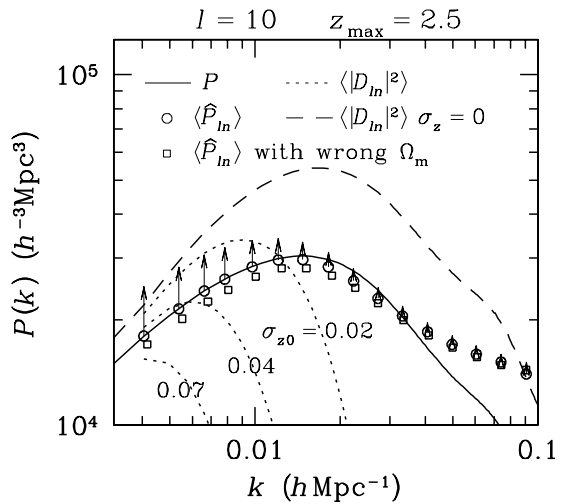}
\caption[f3]{Estimated \thrd{} matter power spectrum 
$\langle \hat{P}_{ln} \rangle$ 
(circles) for $l = 10$ and $z_{\rm max} = 2.5$.
The matter power spectrum is estimated for the case
$\sigma_{z0} = 0.07$, where the rms photometric redshift 
error is parametrized as $\sigma_z = \sigma_{z0} (1+z)$.
The recovery works better with a smaller $\sigma_{z0}$.
The arrows indicate errors of the matter power 
spectrum reconstructed with a galaxy bias that evolves
$30\%$ slower than the fiducial bias model. 
The squares show the estimated matter power 
spectrum under an incorrect assumption of the matter fraction 
$\Omega_{\rm m}=0.3$ instead of the fiducial value 
$\Omega_{\rm m}=0.27$.
The solid line is the real-space matter power spectrum that 
one tries to recover. The dotted 
lines are the redshift-space galaxy power spectrum with the effect
of photometric redshift errors and the linear redshift distortion,
while the dashed line is the redshift-space galaxy power spectrum 
with only the linear redshift distortion.
\label{fig:pk}}
\end{figure}

We calculate the \thrd{} redshift-space galaxy power spectrum 
$\langle |D_{ln}|^2 \rangle = \langle |D_{lmn}|^2 \rangle$ using
equation (\ref{eq:dln}) and recover the real-space matter power
spectrum $\langle \hat{P}_{ln} \rangle$ using the simple estimator
equation (\ref{eq:pln}). The results are shown in Fig.~\ref{fig:pk} 
for $l = 10$ and $z_{\rm max} = 2.5$. The redshift-space galaxy
power spectrum (dotted lines) is strongly suppressed by photometric 
redshift errors at $k \ga H(z) / c \sigma_z$,\footnote{We drop the 
subscripts of the wavenumber for convenience.}
while the simple estimator $\langle \hat{P}_{ln} \rangle$ 
recovers the matter power spectrum reasonably well at 
$k \leq 0.02\,\mpci{}$ assuming an rms photometric 
redshift error of $\sigma_{z0} = 0.04$. The discrepancies between 
the recovered and actual matter power spectra at 
$k > 0.02\,\mpci{}$ are due to 
the broadening of the window function $U_{lnn'}^2$ at high 
wavenumbers and our use of the very simple estimator.  More 
sophisticated estimators can be designed to improve the accuracy of
the recovery. For $\sigma_{z0} = 0.02$, there will be no 
discernible difference between the recovered and the true matter 
power spectra within the range of Fig.~\ref{fig:pk}. 

Since the correction
is so large, one needs to know $\sigma_z$ accurately to 
recover the power spectrum. In the 
(unachievable) case of no redshift errors; i.e., $\sigma_z = 0$, 
$\langle |D_{ln}|^2\rangle$ is boosted relative to $P(k)$ by a factor
that approaches a constant at larger wavenumbers, where the 
plane-parallel approximation is valid and the \citet{k87} formula
applies \citep{ht95}.

\begin{figure}
\centering
\epsscale{1}
\plotone{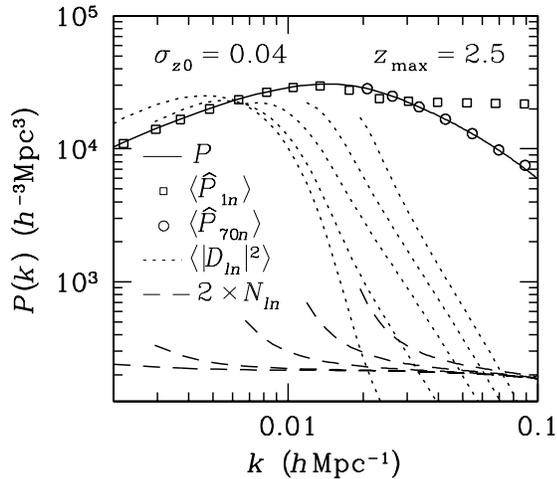}
\caption[f4]{Similar to Fig.~\ref{fig:pk}, but we show the \thrd{} 
redshift-space galaxy power spectrum $\langle |D_{ln}|^2 \rangle$ 
and recovered matter power spectrum 
$\langle \hat{P}_{ln} \rangle$ for different multipoles. 
From left to right, the redshift-space galaxy power spectrum 
(dotted lines) and shot noise $N_{ln}$ (dashed lines) are calculated
with $l = 1,6,20,40$, and $70$. 
The matter power spectrum $\langle \hat{P}_{ln} \rangle$ is 
estimated with $l = 1$ (squares) and $70$ (circles).
The redshift-space galaxy power spectrum
increases with the multipole number on scales dominated by photometric
redshift errors, because at the same 
wavenumber higher multipoles contain a smaller fraction of radial 
modes, which are affected by photometric redshift errors, and a 
larger fraction of angular modes, which are not. 
\label{fig:pkl}}
\end{figure}

Uncertainties of the galaxy bias and cosmological parameters
can cause systematic errors in the estimated matter power spectrum.
The errors arise in two ways: 1) mismatched weight function 
$w(r,k)$ [for measuring the observables, see equation (\ref{eq:Almn})]
and the factor $E_{ln}$ [for recovering the matter power spectrum from 
the observables, see the estimator equation (\ref{eq:pln})],
and 2) wrong length scales in all the measurements and calculations 
if an incorrect cosmological model is assumed.

Current measurements of the galaxy bias at low redshift have 
errors of roughly $10\%$ \citep[e.g.][]{hvg02, verde02, smm05}. If 
the bias is overestimated or underestimated by $10\%$ at all 
redshift, then the normalization of the recovered matter power 
spectrum will be changed by roughly $20\%$ in the opposite direction
without affecting the shape. However, the shape can be altered if 
an error is introduced to the evolution of the bias. The arrows in 
Fig.~\ref{fig:pk} illustrate the distortion of the recovered matter 
power spectrum due to a $30\%$ underestimation of the bias growth 
rate with respect to the fiducial bias model. This case clearly 
shows that understanding the galaxy bias is crucial to a successful
reconstruction of the matter power spectrum from a galaxy survey.

\citet{hj04} demonstrate that with a 4,000 square 
degree next generation ground based weak lensing survey one can 
achieve percent level constraints on the galaxy bias (even as a 
function of scale) by a joint analysis of the shear--shear, 
galaxy--shear, and galaxy--galaxy correlations. Hence, we are 
optimistic that, with the larger LSST survey, the error in the galaxy 
bias will not be the dominant source of error in reconstructing the 
matter power spectrum. 

If one assumes a matter fraction 
$\Omega_{\rm m}=0.3$ instead of the fiducial value 
$\Omega_{\rm m}=0.27$, then the comoving distance to $z=2.5$ will be 
reduced by $3\%$. This leads to roughly a $9\%$ reduction in the raw
galaxy power spectrum and an increase of the wavenumber by $3\%$. We 
recover the fiducial-model matter power spectrum by assuming 
$\Omega_{\rm m}=0.3$. The results are shown as squares in 
Fig.~\ref{fig:pk}. The scaling of the matter power spectrum due to 
the change in distances is visible but somewhat compensated by 
errors in the factor $E_{ln}$ and weight function $w(r,k)$. 
The error in $\Omega_{\rm m}$ does not produce features in the 
recovered matter power spectrum, so that one can 
still detect primordial features on very large scales.

Examples of different multipoles are given in Fig.~\ref{fig:pkl}.
One sees that the redshift-space galaxy power spectrum (dotted lines)
increases with
the multipole number on scales dominated by photometric redshift
errors. The reason is that at the same 
wavenumber higher multipoles contain a smaller fraction of radial 
modes, which are affected by photometric redshift errors, and a 
larger fraction of angular modes, which are not. From the difference
between multipoles of the redshift-space galaxy power spectrum one
may actually quantify the effect of photometric redshift errors 
(and redshift distortions) without knowing $\sigma_z$. 
This is analogous to using the 
quadrupole-to-monopole ratio to determine the redshift distortion
parameter and pairwise velocity dispersion 
\citep[e.g.][]{cole94,pcn01}. The estimated matter power spectrum
$\langle \hat{P}_{ln} \rangle$ with $l = 1$ fails to recover the 
true power spectrum at $k \ga 0.02\,\mpci{}$ because the window 
function $U_{1nn'}^2$ is too broad there to guarantee the accuracy 
of the simple estimator (note that $U_{70nn'}^2$ remains narrow). 
This can be improved with a better estimator.

Fig.~\ref{fig:pkl} also demonstrates that the shot noise $N_{ln}$ is 
indeed negligible on scales of interest for the fiducial survey with 
$z_{\rm max} = 2.5$. Since $\bar{n}_{\rm g}(z)$ increases rapidly with 
decreasing 
redshift, the shot noise for a lower $z_{\rm max}$ is also negligible.
The rise of the shot noise at small wavenumbers is due to the radial 
variation of the selection function as well as the 
wavenumber-dependence of the weight function [see equation 
(\ref{eq:sn})]. Otherwise, $N_{ln}$ would be wavenumber-independent.

Given that the shot noise is very low, we can divide the galaxies into
sub-samples of different luminosity and measure the power 
spectrum of each sub-sample separately. Although sub-sampling cannot 
reduce the cosmic variance, one does gain by reducing the inhomogeneity
in each sub-sample that is used as a (biased) tracer of the cosmic
density field. By comparing the power spectra of  
luminosity classes, we can infer the relative bias between them 
\citep[see e.g.][]{nbh01, zbf02} and test the assumption of 
a scale-independent bias on large scales. 

\begin{figure}
\centering
\epsscale{1}
\plotone{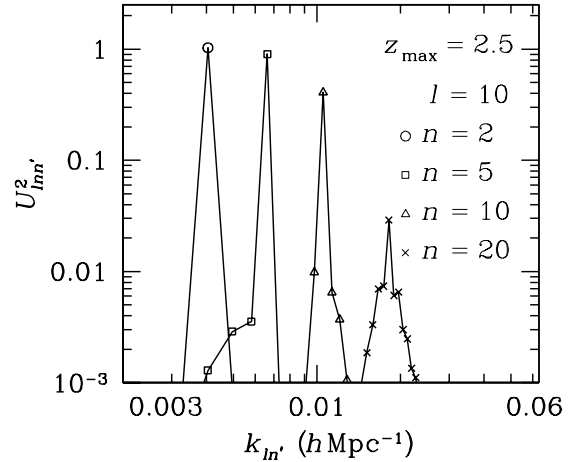}
\caption[f5]{The window functions $U_{lnn'}^2$ that relate the 
\thrd{} real-space matter power spectrum to the redshift-space
galaxy power spectrum [see equation (\ref{eq:dln})]. They are
calculated here with $l = 10$. Symbols
also mark the wavenumbers of several discrete modes $k_{ln'}$. 
They become broader at higher wavenumbers because of photometric
redshift errors and redshift distortion. The window functions 
of other multipoles are similar to those of $l = 10$ but 
shifted toward higher (lower) wavenumbers for higher (lower) 
multipoles.
\label{fig:u2}}
\end{figure}

The estimator $\hat{P}_{ln}$ can be further binned to reduce the
error variance. We calculate the window functions $U_{lnn'}^2$ that 
relate the \thrd{} real-space matter power spectrum to the 
redshift-space galaxy power spectrum using equation (\ref{eq:sn}). 
The $l = 10$ results are shown in Fig.~\ref{fig:u2}. 
Since the window functions are fairly narrow in $k$ space, 
each $\hat{P}_{ln}$  
is nearly independent of each other. Thus, equation (\ref{eq:sigmaP})
is still valid for forecasting the errors, but the mode counting 
becomes
\be \label{eq:NkSHA}
N_k = f_{\rm sky} \sum_{l>0} (2l+1) N_l,
\ee
where $f_{\rm sky}$ is the fraction of the sky covered by the survey, 
and $N_l$ is the number of radial modes of multipole $l$ that fall in 
the band. 
We have neglected the shot noise and approximated the reduction
in the number of modes due to partial sky coverage
by an overall reduction of modes by a factor of
$f_{\rm sky}$. 
For a real survey, one can incorporate the sky cut as well as 
an angular selection function (or mask) numerically. 
We exclude the monopole modes, because they can be confused with the 
radial selection function.
Further modifications to equation (\ref{eq:NkSHA}) are needed in the
regime where the window function $U_{lnn'}^2$ starts to broaden.

After counting spherical harmonic modes, we find that the spherical
geometry only mildly increases the sample variance
of $P(k)$ on the scales of interest, which can be 
identified with the outer error bars in Fig.~\ref{fig:varP}. Hence, 
the simple mode counting with a cubic geometry can still serve as
a reasonable approximation.

\section{Extinction and Photometry Errors} \label{sec:extin}
On very large scales, the variance of density fluctuations in 
logarithmic $k$ bins is very small, 
e.g~$\Delta^2(k) = k^3P(k)/2\pi^2 \sim 10^{-3}$ at $k = 0.01\,\mpci{}$.
An unknown varying extinction over the wide survey area can 
cause fluctuations in galaxy counts that may swamp the signal.
If the logarithmic slope of galaxy counts $\bar{n}_{\rm g}(<m)$ 
as a function of magnitude is
$s = {\rm d}\log \bar{n}_{\rm g}/{\rm d} m$, then the fractional 
error in galaxy counts is
\be
\frac{\delta n_{\rm g}}{n_{\rm g}} = \ln 10\, s\, \delta m
= 2.5\, s \frac{\delta f}{f},
\ee
where $\delta f/f$ is the fractional error in flux caused by, e.g., 
extinction correction residuals. Observationally 
$s$ varies from $0.6$ at blue wavelengths to $0.3$ in the red 
\citep[e.g.][]{t88,pmz98,yfn01}, and tends to be flatter for fainter 
galaxies \citep{msc01,lld03}. To keep this systematic angular 
fluctuation well below $\Delta(k)$, one has to reduce the flux error 
to 1\% or better over the whole survey area. This is a very 
conservative estimate, because the power spectrum receives 
contributions from not only angular clustering but also radial 
clustering of galaxies on large scales, which is much less affected 
by the extinction or photometry errors. 

Extinction $A_\lambda$ (in mag) is related to reddening via color 
excess $E(B-V)$.
Aside from a zero-point difference of $0.02$ mag, the reddening maps 
made with different methods by \citet{bh82} and \citet*{sfd98} agree 
with each other up to an rms error of 0.007 mag \citep{b03}. With 
$A_B = 4.3 E(B-V)$, the relative error in reddening translates 
to an error of 0.03 mag in $B$ magnitude or a flux error of 3\%. 
\ion{H}{1} data and galaxy counts were combined to produce the 
reddening map in \citet{bh82}. At that time galaxy counts were 
typically less than 100 per square degree and thus prone to 
statistical error. With 250,000 galaxies per square degree\footnote{
Since the \twod{} projection of galaxy counts does not require redshift
information, one can use all galaxies that have good photometry.}, 
the LSST will achieve the same accuracy in differential extinction with 
\twod{} galaxy 
counts alone, as long as the dust, or anything that alters the flux, 
is largely confined in our galaxy, or, at least, at very low redshift.

The number of galaxies $N_{\rm g}$ within an angular window 
$\Theta(\hat{\mathbf{r}})$ is given by
\bea
N_{\rm g} &=& \int n_{\rm g}(\mathbf{r}) \Theta(\hat{\mathbf{r}})
{\rm d}^3 r \\ \nonumber
&=& \int \bar{n}_{\rm g}(r)\Theta(\hat{\mathbf{r}})
[\delta_{\rm g}(\mathbf{r}) + 1] {\rm d}^3 r.
\eea
The mean number counts is
\be
\bar{N}_{\rm g} = \int \bar{n}_{\rm g}(r) \Theta(\hat{\mathbf{r}}) 
{\rm d}^3 r,
\ee
and the variance is
\be \label{eq:varN}
\sigma_{N_{\rm g}}^2 = \sum_{lm} \int P(k) |n_l(k) \Theta_{lm}|^2
k^2 {\rm d} k,
\ee
where
\bea \nonumber
n_l(k) &=& \sqrt{\frac{2}{\pi}}\int \bar{n}_{\rm g}(z)b(z)g(z) j_l(kr)
              {\rm d} z, \\ \nonumber
\Theta_{lm} &=& \int \Theta(\hat{\mathbf{r}}) 
Y_{lm}^*(\hat{\mathbf{r}}) {\rm d} \hat{\mathbf{r}}.
\eea
We have used the identities
\bea
&&  \bar{n}_{\rm g}(r) r^2 {\rm d} r =  \bar{n}_{\rm g}(z) {\rm d} z \\
&& e^{i\mathbf{kr}} = 4\pi\sum_{lm} i^l j_l(kr)Y_{lm}(\hat{\mathbf{r}})
Y_{l-m}(\hat{\mathbf{k}})
\eea
to derive equation (\ref{eq:varN}).  
The variance of $N_{\rm g}$ is
analogous to the rms density fluctuation within a volume 
\citep[e.g.][]{zf03}.

For demonstration purpose, we calculate 
the variance with an angular window function
\[
\Theta(\theta, \phi) \equiv \Theta(\theta) = 
e^{- \theta^2/2\theta_{\rm M}^2},
\]
where $\theta$ and $\phi$ are, respectively, the polar and azimuthal 
angles. The multipole number $l$ can be approximately related to the 
characteristic scale of the window function $\theta_{\rm M}$ by 
$l \sim 360^{\rm o} / \theta_{\rm M}$. 

For the effect of the galactic extinction, we use the same axial 
symmetric window function, and place it randomly on the reddening 
map \citep{sfd98} with a restriction that its central galactic 
latitude $|b_{\rm c}| > 20^o + 1.5 \theta_{\rm M}$. 
The rms fluctuation of $B$-band galaxy counts within the window 
function is calculated with the conversion 
$\delta n_{\rm g}/n_{\rm g} \sim \delta A_B = 4.3 \delta E(B-V)$. 

\begin{figure}
\centering
\epsscale{1.1}
\plotone{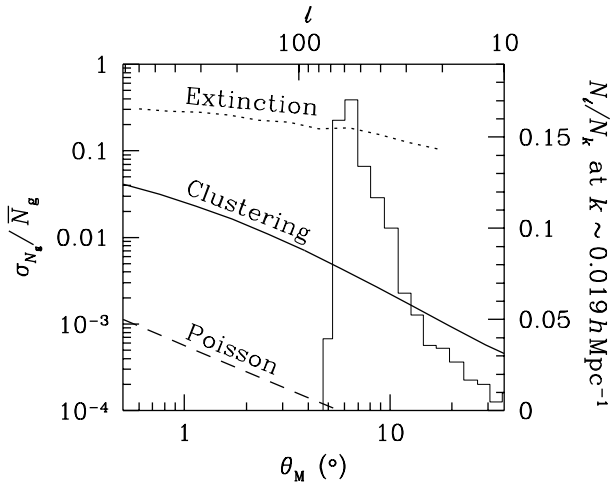}
\caption[f6]{Rms fluctuations of galaxy counts due to the 
extinction (dotted line), galaxy clustering (solid line), and 
Poisson noise $1/\sqrt{\bar{N}_{\rm g}}$ (dashed line) within 
an angular window of size $\theta_{\rm M}$. 
The histogram shows the contribution to the number of independent
modes $N_k$ ($k \sim 0.019\,\mpci{}$) from
each band of multipoles for the highest waveband in 
Fig.~\ref{fig:varP}. The scale of the histogram is marked on the 
right axis. For lower wavebands, the distribution moves 
to lower multipoles. The multipole number $l$ is related to 
$\theta_{\rm M}$ by $l \sim 360^{\rm o} / \theta_{\rm M}$.
\label{fig:varN}}
\end{figure}

The results are shown in Fig.~\ref{fig:varN}. The galactic 
extinction apparently has a dominant effect over galaxy clustering. 
However, the rms fluctuation due to galaxy clustering does not 
include radial clustering because of the projection in redshift 
direction. When the radial information is restored, the clustering 
variance will be much higher. If one restricts the window function 
to higher galactic latitude, the rms fluctuation due to extinction 
will be smaller. In addition, one can also correct for
the fluctuations in the extinction; what will hinder the
measurement of the galaxy power spectrum is rather the 
uncertainties in the extinction.

If one attributes all the angular variations in the projected galaxy 
surface number density to extinction (or any form of photometry 
errors), the rms residual of relative flux error will be set by the 
intrinsic rms fluctuation of the projected galaxy number density due
to clustering, which is well 
below 1\% on scales above several degrees, where the contribution to 
$N_k$ ($k \sim 0.019\,\mpci{}$) peaks. For smaller $k$s (larger 
scales), the contribution comes from lower multipoles, and the 
rms fluctuation of the number counts decreases further.
This means that the relative flux error over 
several square degrees can be controlled to better than 1\% by 
\twod{} galaxy counts from the LSST, which is sufficiently accurate 
for measuring the \thrd{} power spectrum on the largest scales. 

Further improvement on the relative flux error is possible by combining
galaxy counts with multi-band photometry 
\citep*[e.g.][]{bwm05} and \ion{H}{1} and CO surveys. 
Since the Poisson noise in the galaxy counts is an order of magnitude 
lower than that caused by galaxy clustering, one can also divide the 
galaxies into groups of similar properties and compare them in one 
field with those in another to better determine the differential
extinction. 

\section{Applications} \label{sec:app}

Determination of the mass power spectrum has many scientific
applications. They include searching for dark energy fluctuations, 
determining the dark energy equation of state, determining the sum of 
neutrino masses, and probing the primordial power spectrum of 
fluctuations produced during inflation. Here we consider two applications:
determining the primordial power spectrum to probe inflation and 
determining the distance-redshift relation in order to probe dark
energy.

\subsection{The Primordial Power Spectrum of Density Perturbations}
The mass power spectrum on large scales is a direct measure
of the primordial fluctuations, and it provides a means to probe the 
generator of these fluctuations. 
Recent CMB observations have revealed some puzzling properties of the
largest scales
\citep{peiris03, efstathiou04b, schwarz04, deoliveira04, eriksen04b,
hansen04, land05, jaffe05}.
These peculiar features may very well have their origin in systematic 
error. However, they have at the very least served to draw attention to 
the possibility that the very large-scale structure of the Universe 
may be more complex than in our simplest models.

For specificity, let us focus on one of these puzzles and see how
a large-volume photometric redshift survey could shed further light
on the solution.  In \citet{peiris03} it is noted that there are
some unusual sharp features, or `glitches', in the rise to the
first peak of the WMAP temperature power spectrum.  \citet{peiris03}
also suggest that the glitches may be evidence of features in the
primordial power spectrum.  Adjusting two parameters of an inflation
potential model with a sharp break, or `step' \citep*[e.g.][]{gms91}, 
they find a significantly better fit to the power spectrum than if 
they assume a featureless power law.  The step inflation model leads
to a new physical length scale (the size of the Horizon as the inflation
scalar field crosses the feature in its effective potential) that imprints itself on the
power spectrum as a series of peaks and troughs. 

\citet{peiris03} point out that the feature in the primordial power
spectrum is detectable in future large-volume redshift surveys.  One
can indeed see in Fig.~\ref{fig:varP} that the statistical power is there
to map out the wiggles with a high signal-to-noise ratio.  

These features are much more pronounced in the matter power spectrum
than in the temperature power spectrum.  The CMB projection from three
dimensions to two dimensions makes them less prominent.  In general
it is interesting to compare the relative abilities of the CMB and
galaxy surveys to constrain the primordial power spectrum.  

Before turning to forecasts from CMB data though, let us first refine
our forecasts from galaxy survey data at $k > 0.01$ \mpci{}.  As seen 
in Fig.~\ref{fig:pkl}, one has to apply a large correction for
photometric redshift errors and, to a much less extent, redshift
distortions to recover the matter power spectrum at $k \ga 0.01\
\mpci{}$.  An incorrect probability distribution for
the redshift errors can introduce significant errors in the
recovered matter power spectrum.  Even if the correction is accurately
determined, one will still be limited by the shot noise at $k \ga 0.1\
\mpci{}$, which is boosted by the correction along with the observed
galaxy power spectrum.

\begin{figure}
\centering
\epsscale{1}
\plotone{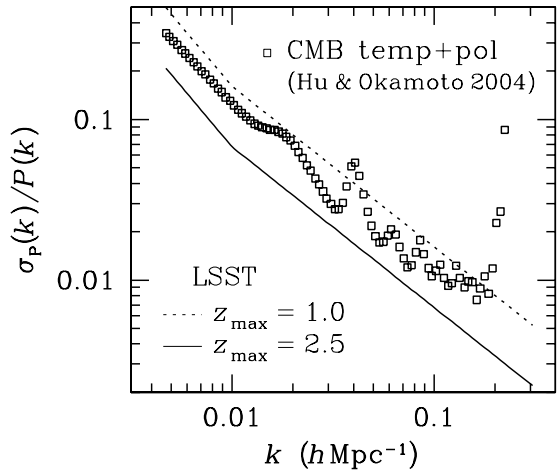}
\caption[f7]{Forecasts of sample variance errors on the primordial 
matter power spectrum. For the LSST, we set $z_{\rm max} = 1.0$ 
(dotted line) and $2.5$ (solid line), and the mode counting follows 
equation (\ref{eq:Nk2}) with $k_{\|}^* = 0.01\,\mpci{}$.
The forecast for the CMB (open squares) includes both temperature 
and polarization information, and it is taken from \citet{ho04}. 
Both forecasts assume a binning of $\Delta k = 0.05k$.
\label{fig:compv}}
\end{figure}

To avoid sensitivity to uncertain and large corrections for suppression
of power due to photometric redshift errors, we can simply discard
modes with a lot of suppression in the radial direction.  That is,
as discussed in \citet{se03} and \citet{gb05}, we can discard modes with
$k_\|$ above some critical value.  The approximate
mode counting equation (\ref{eq:Nk}) becomes
\be  \label{eq:Nk2}
N_k = \left\{ \begin{array}{ll} 
k^2 \Delta k V / 2\pi^2      & \quad k_\| \leq k_\|^* \\
k k_\|^* \Delta k V / 2\pi^2 & \quad k_\| >    k_\|^*, \\
\end{array} \right.
\ee
where a cubic geometry and the plane-parallel approximation are 
assumed.  One would choose $k_\|^*$ to be small enough so 
that residual errors in the power 
spectrum can be tolerated after correcting for photometric 
redshift errors and redshift distortion. We set 
$k_\|^* = 0.01\,\mpci{}$ and 
assume that the residual error is negligible for $k_\| \leq k_\|^*$.

Note that at low $k$ the number of modes scales with $k$ as a 
three-dimensional
survey, increasing as $k^3$ if $\Delta k \propto k$, 
and that above $k_\|^*$ the scaling is 
that of a two-dimensional
survey.  Thus, although the CMB will always be the best on length scales
larger than those that fit in the galaxy survey, a galaxy survey can
compete, at least in terms of the statistical weight, with the 
CMB on sufficiently small scales.  If the cross over scale is large
enough, galaxy bias modeling will be simple enough to exploit this
statistical improvement.  

We present in Fig.~\ref{fig:compv} the sample variance errors of the 
primordial
matter power spectrum for the LSST with $z_{\rm max} = 1.0$ (dotted 
line) and $2.5$ (solid line). Also included is the error forecast for 
the CMB (open squares) using both temperature and polarization 
information \citep{ho04}. It is very encouraging that the LSST can 
almost match the best-case-scenario CMB results with $z_{\rm max} = 1.0$
and even provide a better measurement of the primordial power spectrum
with $z_{\rm max} = 2.5$. The cross over scale referred to in the
previous paragraph is indeed at very large scales.  For the 
$z_{\rm max}=2.5$
case, it is at $k \lesssim 5 \times 10^{-3}\,\mpci{}$.

The binning in Fig.~\ref{fig:compv} is for $\Delta k = 0.05k$
for ease of comparison with Hu \& Okamoto's results.  However, 
the spectral resolution of the galaxy survey is
$7.5\times10^{-4}\,\mpci{}$ ($z_{\rm max} = 2.5$) with exceptions for 
the first few modes of each multipole, which have slightly coarser 
resolutions.  Thus for $k < 0.015\,\mpci{}$ the binning assumed in 
Fig.~\ref{fig:compv} is too fine.  To avoid correlations, that would
otherwise increase the errors above the values plotted in the figure,
one could bin more coarsely.  This coarser binning would {\em reduce}
the level of error below what is plotted in the figure by 
a factor of $\sqrt{0.05 k/\Delta k}$, although the
number of points would, of course, be reduced also.  

\citet{ho04} point out that the CMB power spectrum errors in
Fig.~\ref{fig:compv} are highly correlated and that there are certain
linear combinations of the power spectrum measurements with much
smaller errors.  Approximately 50 principal-component modes of the
covariance matrix can be measured to percent level precision.  The
best-determined principal-component modes are combinations of Fourier
powers at $k \ga 0.08\,\mpci{}$.

Our forecast for how well the
primordial power spectrum can be determined from galaxy survey data
is done in the limit of no uncertainty in $r(z)$, $g(z)$, $b(z)$,
$\omega_{\rm m}$ and $\omega_{\rm b}$.  We expect that CMB data from \emph{Planck} 
and the galaxy
survey data itself can be used to reconstruct these well enough that
the uncertainties do not qualitatively change our error forecasts.
The \citet{ho04} forecast does include the effect of these uncertainties
which are more important for the CMB because of the projection from
three dimensions to two dimensions and the greater prominence of the
acoustic oscillation features.

We must mention two further caveats.  First, for the lowest $k$ 
values the spherical geometry slightly increases the variance beyond 
what one gets for a cubic geometry as shown in Fig.~\ref{fig:varP}. 
Second, whether one can neglect errors in the correction for 
photometric redshift power suppression at $k < 0.01\,\mpci{}$ remains 
to be demonstrated.

\subsection{Standard Rulers For Determining $r(z)$}
In the previous subsection we considered how well the primordial
power spectrum could be determined with $r(z)$, $g(z)$, $b(z)$, 
$\omega_{\rm b}$ and $\omega_{\rm m}$ known perfectly.  Now we 
consider how well $r(z)$ can be determined if we
assume the primordial power spectrum to be featureless, as expected
in the simplest models of inflation.  If the power spectrum
remained featureless, it would be impossible to reconstruct
$r(z)$ from galaxy clustering data, but features imprinted by
the evolution of this spectrum can be used as standard rulers
for determination of the angular-diameter distance.  

In linear perturbation theory there are two length scales that become
imprinted on the matter power spectrum.  The larger scale feature is
the peak near $k = 0.02\,\mpci{}$ that can be seen in
Fig.~\ref{fig:pk}.  The location of this feature depends on the size
of the horizon at the epoch of matter-radiation equality.  Assuming a
standard radiation content, this is set by the matter density,
$\omega_{\rm m}$, which can be determined to sub-percent accuracy with
\emph{Planck} \citep*{eisenstein99}.  The smaller scale feature is a 
series of peaks and troughs in the matter power spectrum due to acoustic
oscillations the baryons undergo in the pre-recombination plasma
\citep{py70, be84, h89, hs96}.  The
length scale here is the comoving size of the sound horizon at
last-scattering.  This depends on $\omega_{\rm m}$ and $\omega_{\rm b}$ 
which can
both be determined well from CMB observations.  Thus, both of these
features can serve as CMB-calibrated standard rulers,
which one can use to determine the comoving angular-diameter distance,
$r(z)$ \citep{eht98, cooray01}.

Since $r(z) = c\int_0^z dz'/H(z')$ these measurements can be used to map
out the history of the expansion rate and thereby provide constraints
on the dark energy.  

First we will consider the baryonic oscillations.  
The lowest $k$ peak in the series of baryonic oscillations
has already been observed
in spectroscopic redshift surveys 
\citep{ezh05, cpp05}.  Several papers have studied
how well $r(z)$ and cosmological parameters can be determined from 
future surveys, both
spectroscopic and photometric \citep*{bg03, hh03, l03, se03, bb04, 
djt04, abf05, gb05, linder05, zhan05}.  Prospects for controlling non-linear
evolution and galaxy biasing look very good \citep{white05, se05}.
Photometric surveys over larger volumes can provide 
constraints that are competitive with spectroscopic surveys over
smaller volumes.

If one counts the modes using equation (\ref{eq:Nk2}), the shot noise
will continue to be negligible at $k > 0.1\,\mpci{}$ and the dominant
errors there will come from the uncertainties of the correction for 
photometric redshift errors, nonlinear redshift distortion, nonlinear 
evolution, and galaxy bias. Despite the complexities, these errors 
do not produce oscillating features in the power spectrum. Thus it is 
possible to use a photometric redshift survey to measure the 
angular-diameter distance accurately.

We now turn to a discussion of the large-scale feature.  
Current galaxy surveys cannot be used to measure the power spectrum on 
scales larger than $k = 0.02\,\mpci{}$ due to their limited survey
volume. The LSST, on the other hand, will be able to probe scales with
wavenumbers as small as several thousandths \mpci{}. 
At the expense of increased 
sample variance error, one may measure the matter power spectrum in 
several redshift bins with the LSST. For example, Fig.~\ref{fig:compv} 
suggests that 4 equal-volume bins from $z = 0$ to $2.5$ will enable us
to measure the matter power spectrum to roughly 10\% around 
$k = 0.01\,\mpci{}$.  The angular scale to which this feature projects
depends on the angular-diameter distance to each redshift shell, 
enabling one to determine the angular-diameter distance.  

Generally, constraints from the large-scale feature will be weaker
than from the baryonic oscillations for two reasons:  first, the latter 
are much sharper and therefore less tolerant of the horizontal shifts 
induced by a re-scaling of distances and second, the sample variance
errors are larger on larger scales.  

However, even this broad feature can lead to powerful 
distance determinations.  \citet{se03} find that the angular-diameter
distance determinations using the large-scale feature have 2 to 3
times larger errors.  Thus, though subdominant in their constraining
power, the large scale feature offers a useful independent check on
distance determination from the baryonic oscillations.

With the LSST and CMB priors from the \emph{Planck} mission, 
one can achieve errors of $\sim 1\%$ on $r(z)$, $0.10$ on $w_0$, and 
$0.25$ on $w_{\rm a}$ using both the baryon oscillations and 
broadband feature \citep{zhan05},
where the dark energy equation of state is parametrized as 
$w(z) = w_0 + w_{\rm a}[1-(1+z)^{-1}]$.

\section{Discussion and Conclusions} \label{sec:dis}

For the reconstruction of the matter power spectrum $P(k)$ to work, 
we must know the comoving distance $r(z)$, the linear growth 
function $g(z)$, the galaxy bias $b(z)$, and the rms 
photometric redshift error $\sigma_z$ adequately well. 

One may improve the knowledge of the galaxy bias on large scales
by comparing galaxy statistics with mass 
statistics inferred from weak lensing (\citealt{hvg02}; 
\citealt{smm05}; for a joint analysis, see \citealt{hj04}), by 
constructing realistic theoretical models \citep[e.g.][]{ps00}, 
and by studying galaxies in large volume hydrodynamical simulations 
that include star formation and feedback \citep[e.g.][]{wdk04}. 
Available observational constraints on bias \citep[e.g.][]{tbs04}
constrain it on scales smaller than those of interest here,
but since the LSST is designed for weak lensing, 
it can help to measure the galaxy bias on large scales for 
galaxies associated with the lensing mass field.

The uncertainties in $r(z)$ and $g(z)$ can be mitigated by 
assuming a prior for the cosmological model, but at the same time we 
lose the ability to measure the matter power spectrum 
model-independently. However, since cosmological information is 
encoded in the matter power spectrum through multiple channels in 
addition to $r(z)$ and $g(z)$, cosmological parameters can be 
estimated from the reconstructed matter power spectrum as well.
This will provide a self-consistency check for the choice
of the parameters.

Photometric redshift errors strongly suppress the raw galaxy 
power spectrum $\langle |D_{ln}|^2 \rangle$. 
Thus, it is crucial to determine the correction 
for photometric redshift errors. We have adopted a form of this
correction from \citet{ht95}, but other forms also exist
\citep[e.g.][]{ballinger96}. They should be tested against $N$-body 
simulations to improve the accuracy. Meanwhile, it appears possible 
from Fig.~\ref{fig:pkl} that one can infer the correction by 
comparing the galaxy power spectrum of different multipoles. One may
also discard high $k_\|$ modes to subdue the errors due to 
uncertainties in the correction (see Fig.~\ref{fig:compv}).

The extinction is yet another challenge to measuring the matter power
spectrum on very large scales. We have demonstrated that with 250,000
projected galaxies per square degree the LSST can actually determine 
the differential extinction to better than 1\% from galaxy counts 
alone. This means that not only can we be sure that the signal is truly 
coming from clustering of galaxies but we can also combine the galaxy counts 
with multi-band photometry \citep[e.g.][]{bwm05} 
and \ion{H}{1} and CO maps to produce a 
more accurate reddening map, which will be useful for other observations
such as CMB surveys.

Despite all the difficulties, we find that a deep and wide photometric 
redshift survey, such as can be done with the LSST, can measure the power 
spectrum of primordial fluctuations on very large scales. The 
shot noise on these scales is so low (see Fig.~\ref{fig:pkl}), that it 
is possible and beneficial to divide the galaxies into different types 
to check for consistency and to indirectly test the 
scale-independence of the galaxy bias on large scales.

The large-scale primordial power spectrum measured from galaxies will 
complement the CMB experiments and provide valuable insights on 
inflation. For example, it can be used to examine the possibility of 
a step inflation potential \citep{peiris03}. The large-scale peak and 
intermediate-scale baryon oscillations in the matter power spectrum 
will provide useful constraints on cosmological parameters including 
the dark energy equation of state parameters. 
They can also reduce the uncertainties of the photometric 
redshift bias \citep{zhan05}, and thus prevent 
the redshift bias from severely degrading the constraining 
power of the weak lensing tomography on the dark energy equation of 
state parameters (\citealt{huterer05}; \citealt*{mhh05}).


\acknowledgements
We thank D.~Burstein, A.~Connolly, D.~Eisenstein, and L.~Hui for useful 
conversations about reddening maps, photometric redshift errors, and 
baryon acoustic oscillations. We also thank the referee for 
helpful comments.
This work was supported by the National Science Foundation under 
Grant No. 0307961 and 0441072 and NASA under grant No. NAG5-11098.

\appendix  
\section{Spherical Harmonic Analysis} \label{sec:appendix}
For a survey that covers a large fraction of the sky, the spherical 
coordinate system is a more suitable basis to work with. The spherical
harmonic analysis has been applied to the \emph{IRAS} galaxy catalog 
(\citealt{shl92}; \citealt*{fsl94}), the PSCz galaxy catalog 
\citep{tbt99}, and, 
recently, the 2dFGRS catalog \citep{pbh04}. We adopt the formulation in 
\citet{ht95} and describe it here briefly.

The spherical harmonic decomposition of a quantity $A(\mathbf{r})$ is
\be \label{eq:Almn}
A_{lmn} = c_{ln}\int_0^{r_{\rm max}} w(r,k_{ln}) A(\mathbf{r}) 
j_l(k_{ln}r) Y_{lm}^*(\hat{\mathbf{r}}) {\rm d}^3r,
\ee
where an extra weight function $w(r,k_{ln})$ is inserted to minimize the 
variance in the power spectrum measurement. The discrete wavenumber 
$k_{ln}$ is the $n$th root of the boundary condition 
${\rm d} j_l(kr) / {\rm d} r |_{r_{\rm max}} = 0$. The normalization 
$c_{ln}$ satisfies $c_{ln}^{-2} = \int j_l^2(k_{ln}r) r^2 {\rm d}r$.
One can relate the observed \thrd{} galaxy overdensity $D_{lmn}$ to the 
underlying mass overdensity $\delta_{lmn}$ through 
\bea \label{eq:Dlmn}
D_{lmn} &=& \sum_{n'n''}S_{lnn'} Q_{ln'n''} \delta_{lmn''}, \\ 
S_{lnn'} &=& c_{ln} c_{ln'} 
\int \frac{e^{-(r-y)^2/(2\sigma_z^2)}}{\sqrt{2\pi} \sigma_z} 
j_l(k_{ln}r) j_l(k_{ln'}y) r {\rm d}r\, y {\rm d}y, \\ \label{eq:Qlnn}
Q_{lnn'} &=& c_{ln} c_{ln'} \int \bar{n}_{\rm g}(r) g(z) 
\Big\{ b(z) w(r,k) j_l(k_{ln'}r) j_l(k_{ln}r) + \nonumber \\
&& \qquad\qquad\qquad\qquad\quad
\frac{\Omega_{\rm m}^{0.6}(z)}{k_{ln'}^2} 
\frac{\rm d}{{\rm d}r}[w(r,k) j_l(k_{ln}r)]
\frac{\rm d}{{\rm d}r}j_l(k_{ln'}r)\Big\}r^2 {\rm d}r,
\eea
where $\sigma_z$ is the photometric redshift error in comoving distance
units, $\Omega_{\rm m}(z)$ is the ratio between the cosmic matter 
density to the critical density at redshift $z$, and we have applied the
approximation ${\rm d}\ln g / {\rm d}\ln a \simeq \Omega_{\rm m}^{0.6}$ 
\citep{llp91}. For simplicity, we assume a full-sky survey with no angular 
variation in the selection function, so that the galaxy distribution
$n_{\rm g}(\mathbf{r}) = \bar{n}_{\rm g}(r)[b(z) g(z) \delta(\mathbf{r})+1]$,
in which $\delta(\mathbf{r})$ is scaled to $z = 0$. 
The term $S_{lnn'}$ 
would account for the nonlinear redshift distortion, i.e.~the pair-wise 
velocity dispersion, if there were no error in redshift measurements. 
Since the photometric redshift error 
is far greater than the pair-wise velocity dispersion, the latter is 
neglected. The term $Q_{lnn'}$ is due to the linear redshift distortion
\citep[see also][]{s04}.
With $\langle \delta_{lmn} \delta_{l'm'n'}^* \rangle = 
P(k_{ln}) \delta_{ll'}^{\rm K} \delta_{mm'}^{\rm K} 
\delta_{nn'}^{\rm K}$, one can show that
\be \label{eq:dln}
\langle |D_{ln}|^2 \rangle = \langle |D_{lmn}|^2 \rangle 
= \sum_{n'} U_{lnn'}^2 P(k_{ln'}) + N_{ln},
\ee
where the window function $U_{lnn'}^2$ and shot noise $N_{ln}$, 
respectively, are given by
\be \label{eq:sn}
U_{lnn'} = \sum_{n''} S_{lnn''}Q_{ln''n'}\quad\mbox{and}\quad
N_{ln} = c_{ln}^2\int \bar{n}_{\rm g}(r)w^2(r,k)j_l^2(k_{ln}r) r^2{\rm d}r.
\ee

The weight function $w(r,k)$ is introduced in equation (\ref{eq:Almn}) to
reduce the variance in the power spectrum estimator \citep{fkp94,ht95}. 
Designing an optimal weight function is beyond the 
scope of this work. Here, we only demonstrate the performance of 
the estimator with a weight function
$w(r,k) = P(k)/[1 + \bar{n}_{\rm g}(r) g(z) b(z) P(k)]$
\citep[see also][]{fkp94,ht95}.
Fig.~\ref{fig:u2} shows several examples of the window function 
$U_{lnn'}^2$ . The fiducial cosmological 
parameters are given by the \emph{WMAP} results \citep{svp03}, and the 
matter power spectrum is calculated using the fitting formula from 
\citet{eh99}. We model the evolution of an overall bias 
with $b(z) = 0.84 z + 1$ \citep[e.g.][]{wdk04}. The symbols mark the 
discrete wavenumbers of the radial modes. The window function 
$U_{lnn'}^2$ is fairly narrow for $k < 0.02\,\mpci{}$, except
for $U_{l1n'}^2$ (not shown), which are discarded in the mode
counting. The results of other multipoles share 
the same characteristics, except that for higher multipoles the modes 
move toward large wavenumbers and vice versa.


\end{document}